\documentclass[a4paper]{aa}
\usepackage{psfig}
\usepackage{epsfig}
\usepackage{graphicx}

\def \xmm {XMM--Newton}
\def \sax {BeppoSAX}
\def \src {GRS\thinspace1758--258}

\def \nh {N${\rm _H}$}

\def \ferg {erg cm$^{-2}$ s$^{-1}$}
\def \hcm {\hbox {\ifmmode $ atom cm$^{-2}\else atom cm$^{-2}$\fi}}
\def \arcmin {\hbox{$^\prime$}}

\def \chisq {$\chi ^{2}$}

\def\approxgt{\mathrel{\hbox{\rlap{\lower.55ex \hbox {$\sim$}}
        \kern-.3em \raise.4ex \hbox{$>$}}}}
\def\approxlt{\mathrel{\hbox{\rlap{\lower.55ex \hbox {$\sim$}}
        \kern-.3em \raise.4ex \hbox{$<$}}}}

\newcommand {\Msun}{M_\odot}
\begin{document}

%\thesaurus{(02.01.2; 08.09.2; 08.14.1; 10.07.3; 13.25.2; 13.25.3)}

\title{The broad band X--ray spectrum of the black hole candidate \src}

\author{L. Sidoli
%\inst{1}
        \and S. Mereghetti
%        \and et al.\inst{1}
}
\offprints{L.Sidoli (sidoli@ifctr.mi.cnr.it)}

\institute{
        Istituto di Astrofisica Spaziale e Fisica Cosmica --
        Sezione di Milano ``G. Occhialini" -- CNR \\
         via Bassini 15, I-20133 Milano, Italy}

\date{Received 11 February 2002; Accepted: 8 April 2002 }

\authorrunning{L. Sidoli \& S. Mereghetti}

\titlerunning{{\sax\ observation of \src}}

\abstract{We present the results of a  \sax\ observation of the
black hole candidate \src\ carried out in 1997, while the
source was in its low/hard state.
The X--ray  spectrum, simultaneously observed over the broad energy range
from  0.1 to 200~keV,
can be well described by a Comptonized emission model with
electron temperature kT$_{\rm e} = 31.4 ^{+3.4} _{-2.5}$ keV
and  optical depth
$\tau = 4.0 ^{+0.2}_{-0.3}$ (spherical geometry), 
although a cut-off power-law and a reflection model cannot be excluded. 
Additionally, a broad iron line at 6.4~keV with equivalent width  
$EW = 67 ^{+80} _{-40}$~eV has been marginally
detected.
The 0.1--200~keV luminosity is $1.4\times10^{37}$~erg~s$^{-1}$
for an assumed distance of 8~kpc.
The soft and hard luminosities are such that the source falls
inside the so-called ``burster box".
No evidence for a soft excess is present.
}
\maketitle

\section{Introduction}
\label{sect:intro}

The hard X--ray source \src\
was discovered in 1990 (Mandrou \cite{m:90}; Sunyaev et al. \cite{s:91})
during the observations of the Galactic Center region performed
with the GRANAT satellite.
This source and the more famous  ``microquasar'' 1E~1740.7--2942
(e.g. Churazov et al. \cite{c:94})
were found to be the only
persistent sources of hard X-rays (E$>$40 keV) present within
a few degrees from the Galactic Center direction.

Also \src\ can be considered a  ``microquasar'' due to
its association with a radio source with double-sided
relativistic jets (Rodriguez et al. \cite{r:92}).
Recent $Chandra$ observations determined the X--ray
position of \src\ with a
sub-arcsecond error radius,
confirming the association with the radio point source at the
center of the two radio jets
(Heindl \& Smith \cite{hs:01}).

Monitoring in the X--ray band with the \textit{RossiXTE} satellite
revealed a periodicity of 18.45$\pm{0.03}$~days (Smith et al. \cite{s:00}).
Two possible IR candidate counterparts were
found by  Mart\'\i~et al. (\cite{m:98}) compatible with the position of
the central radio source. However,
Eikenberry et al. (\cite{e:01}), based on  a different astrometry to link the
radio and IR data sets, excluded the objects proposed by Mart\'\i~et al. (\cite{m:98})
and found that there are no stars brighter than
K$_{S}$=20.3 compatible with the radio position
(K$_{S}$ is centered at $2.15~\mu$m).
If confirmed, this strong limit on the IR emission implies a low mass companion star
that makes unlikely the interpretation of the above periodicity as an orbital period.

Although a dynamical mass measurement is not available,
\src\ is considered a black hole candidate on the basis of its hard
emission extending above 100 keV and its spectral similarities with Cyg X-1.
Most of the X--ray observations of \src\ determined  a power-law spectrum
(photon index $\Gamma\sim$1.7--2) with a cut-off above $\sim$30 keV
(e.g. Churazov et al. \cite{c:94}; Keck et al. \cite{k:01}),
typical of the so called low/hard state of black hole candidates.

Here we report the results of the first   observation of \src\ covering
simultaneously  the broad energy range from 0.1 up to 200 keV.
The wide spectral coverage achieved with \sax\ is especially important when searching
for the presence of a soft excess.

\section{Observations and Data Reduction}
\label{sect:obs}

\src\ was observed with \sax\
from 1997 April 10 15:20 to April 11 05:09 UTC.
Here we report the results from the  Low-Energy Concentrator Spectrometer (LECS;
0.1--10~keV; Parmar et al. \cite{p:97}), the Medium-Energy Concentrator
Spectrometer (MECS; 1.8--10~keV; Boella et al. \cite{b:97}),
and the Phoswich Detection System (PDS; 15--300~keV;
Frontera et al. \cite{f:97}) instruments.

\begin{figure}[!ht]
\centerline{\psfig{figure=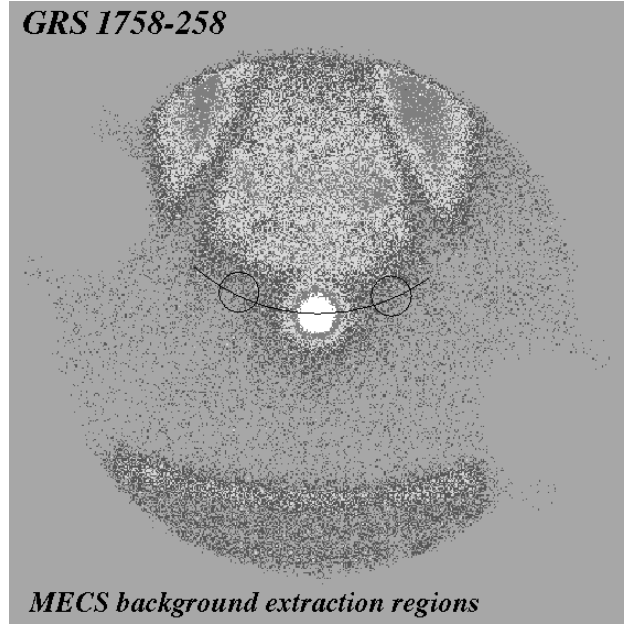,height=75mm,bbllx=70mm,bblly=80mm,bburx=140mm,bbury=250mm}}
\vskip 1.0truecm
\caption{{\small Image of GRS~1758-258 in the 2--10 keV energy range
obtained with the MECS3 unit.
A logarithmic scale has been used to better illustrate the faint contamination
from the straight light of GX 5--1.
The two  MECS background extraction
regions (R=2$'$) near to the source are shown (see text)}}
\label{1758bg}
\end{figure}

The net exposure times
in the LECS, MECS, and PDS instruments were 9.0~ks, 28.3~ks,
and 12.7~ks, respectively.
For   this   observation all the three MECS units were available.
The MECS and LECS field of view (see Fig.~\ref{1758bg} for the MECS)
was contaminated by the stray light
contribution from  the  bright Low Mass X--ray Binary  GX~5--1,
located about 40$'$  away from   GRS~1758--258.
In order to minimize this contamination, the \src\ counts for the
spectral analysis have been extracted from circular regions with radius
(2$'$ for the  MECS and 4$'$ for the LECS)
smaller than that used in standard analysis.
We used the appropriate response matrices that take into account
the size of the extraction region.
To estimate the background for the spectral analysis, we used an \textit{ad hoc} procedure
in order to properly correct  for a background as similar as possible
to that induced by the stray light: the regular shape and
the radial symmetry of the stray light contamination led us to consider
the two circular regions with
radius 2$'$   shown in Fig.~\ref{1758bg}.
Both regions have practically the same spectrum that
represents well the contribution from the GX~5--1  at the position of GRS~1758--258.
We verified that similar results in the fitting parameters (within the 90\% confidence range)
were found also
using a different choice for the LECS and MECS background regions (a standard
annular region around the source position).
This indicates that the background is not critical for this bright source.

The non-imaging
PDS instrument consists of four independent units arranged in pairs, each having a
separate collimator. Each collimator was alternatively
rocked on- and 210\arcmin\ off-source every 96~s during the observation.
These two off-source fields, free from contaminating sources,
were used for the PDS background subtraction.
However, the PDS on--source data were affected by the  GX~5--1 emission.
This effect was relevant only at low energies, due to the
very soft spectrum of GX~5--1 (Gilfanov et al. \cite{g:93}, see their Fig.~6).
We therefore neglected the PDS channels from 15 to 40 keV
in the spectral analysis.
For the same reason,  the non--imaging  data from the HPGSPC instrument,
covering the intermediate energy range between MECS and PDS, could not be used
for the spectral analysis and will not be considered here.

%-------------------------------------------------------------------------------------------------
%-------------------------------------------------------------------------------------------------

\section{Results}

The broad-band spectrum of \src\ was investigated by simultaneously
fitting data from the LECS, MECS and PDS.
The data were selected in the energy ranges
0.1--4.0~keV (LECS), 1.8--10~keV (MECS),
and 40--200~keV (PDS) and rebinned using standard procedures.
The resulting  background-subtracted count rates were
0.74, 3.5, and 3.0~counts s$^{-1}$
for the LECS, MECS, and PDS, respectively.
In the spectral fitting, constant factors were included
to account  for normalization
uncertainties between the instruments.
These factors were constrained
to be within their usual ranges during the fitting.
All spectral uncertainties
and upper-limits are given at 90\% confidence.
Spectral analysis has been performed with {\sc xspec} v.11
software package. 

\begin{figure*}[!ht]
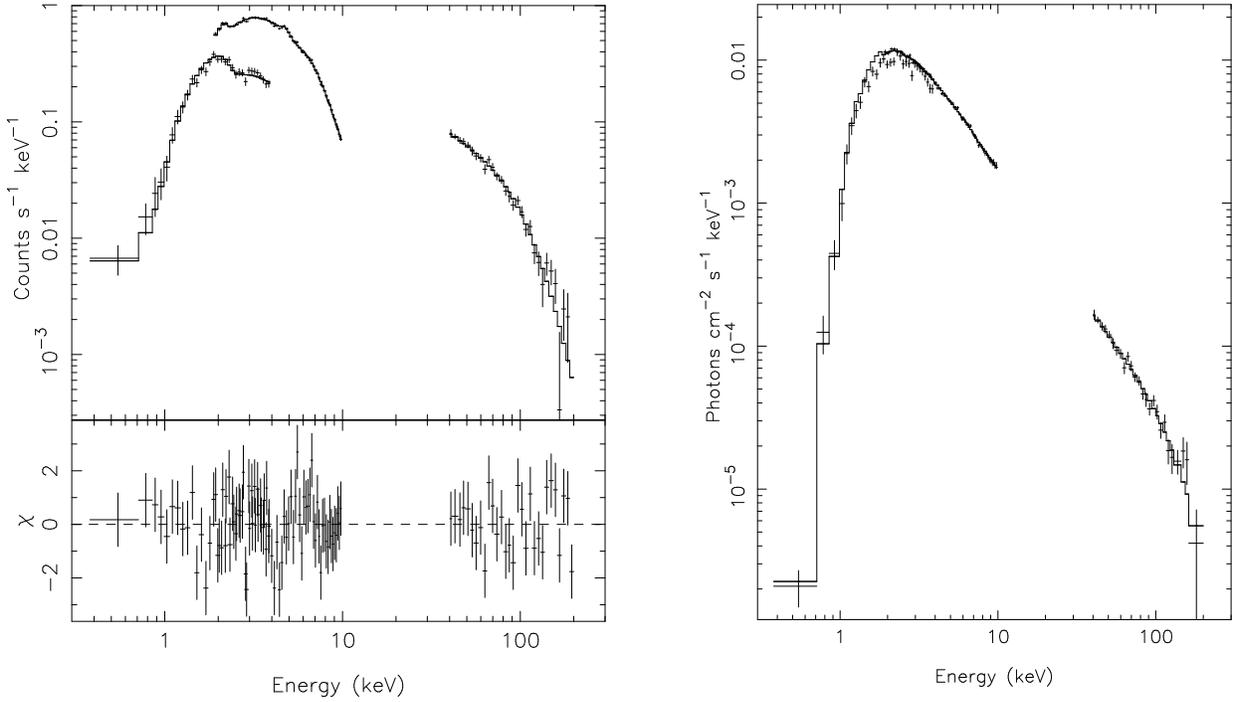

\hbox{\hspace{1.0cm}
\includegraphics[height=8.0cm,angle=-90]{h3485f2a.ps}
\hspace{1.0cm}
\includegraphics[height=7.103cm,angle=-90]{h3485f2b.ps}}
\caption[]{The 0.1--200~keV \sax\ spectrum of \src, fitted with a Comptonization
model (see Table~\ref{tab:spec} for the parameters).
The left panels show the best-fit count spectrum and the
residuals, in units of standard deviations. The right panel shows the
photon spectrum.
}
\label{fig:spec}
\end{figure*}

Initially, an absorbed power-law model was tried,
resulting in photon index $\Gamma$=1.66 and  column density
N$_{\rm H}$=1.85$\times$$10^{22}$~cm$^{-2}$
(\chisq=150.3  for 111 degrees of freedom, dof).
Inspection of the residuals showed that a cut-off in the spectrum
is needed above $\sim$100~keV.
The  inclusion of a high energy cut-off in the power-law model
resulted in a significantly better fit (\chisq/dof=123.7/109),
with the following parameters:
N$_{\rm H}$=$(1.81 ^{+0.08} _{-0.05})$$\times10^{22}$~cm$^{-2}$,
$\Gamma$=1.65$\pm{0.02}$,
cut-off energy E$_{\rm c}$=73$\pm$20~keV
and  e-folding energy E$_{\rm fold}$=$180 ^{+120} _{-80}$~keV.
The  flux in the   0.1--200~keV range, corrected for the absorption,
is $1.95\times10^{-9}$~erg~cm$^{-2}$~s$^{-1}$, which corresponds to
an X--ray luminosity of $1.4\times10^{37}$~erg~s$^{-1}$ (at 8~kpc).
About 33\% of this unabsorbed flux is emitted in the 0.1--10~keV band.
In order to have a more physical picture of the emission spectrum,
we also fitted  a
Comptonization model
({\sc compst} in  {\sc xspec}; Sunyaev \& Titarchuk \cite{st:80}), for which we obtained
a similar good statistical result (\chisq/dof=125.3/110).
The resulting parameters are listed in Table~\ref{tab:spec} and the spectrum
is shown in Fig.~\ref{fig:spec}.

Another model  often used to fit
the low/hard state spectrum of black-hole candidates is
the Comptonization of soft photons in a
hot plasma which includes relativistic effects ({\sc comptt} model in {\sc xspec},
Titarchuk \cite{t:94}).
The best fit with this model gave a slightly worse \chisq\ value than the
{\sc compst} model with the parameters reported in Table 1.

%--------------------------------------------------------
\begin{table}
\begin{center}
\caption[]{Best-fit parameters for the broad-band BeppoSAX
spectrum of \src\ fitted with Comptonization models.
Flux is in the 0.1--200~keV energy range. The luminosity (0.1--200)
has been corrected for interstellar absorption and is for an assumed
distance of   8~kpc}
\begin{tabular}{lll}
\hline
\noalign {\smallskip}
Parameter & COMPST & COMPTT \\
\hline
\noalign {\smallskip}
${\rm N_{\rm H}}$ $(10^{22}$ cm$^{-2}$)  &  $1.83 \pm{0.07}$       &  $(1.4 ^{+0.45} _{-0.21})$\\
kT$_{{\rm e}}$ (keV)                     &  $31.4 ^{+3.4} _{-2.5}$ &  $44^{+146} _{-7}$ \\
$\tau$                                   & $ 4.0 ^{+0.2} _{-0.3} $ &  $3.6^{+0.4} _{-2.3}$  \\
kT$_0$        (keV)                      &          --        &     $<$0.48     \\
Flux (${\rm erg~cm^{-2}~s^{-1}})$        &  $ 2.0 \times 10^{-9}$   &   $ 1.8 \times 10^{-9}$ \\
Luminosity (${\rm erg~s^{-1}})$          &   $ 1.4\times 10^{37}$    & $ 1.3\times 10^{37}$ \\
$\chi ^2$/dof                            &  125.3/110                &  128.3/109 \\
\noalign {\smallskip}
\hline
\label{tab:spec}
\end{tabular}
\end{center}
\end{table}
%----------------------------------------------------------------

%----------------------FIG: COMPST SPECTRUM-------------------------------------
%-------------------------------------------------------------------------------

The addition of  a Gaussian line (centroid energy fixed at 6.4~keV)
to the Comptonization best-fit
resulted in a line width $\sigma$=700 $^{+600} _{-440}$~eV
and in an equivalent width, EW=67 $^{+80} _{-40}$~eV.
The parameters of the continuum remained consistent with those
reported in Table~\ref{tab:spec}
and the \chisq\ value decreased to 116.2 for 108 dof.
An  F-test  indicates that the probability  of such a decrease
occurring by chance is 1.7\%.
Fixing the energy of the line at 6.7~keV   
resulted in a \chisq\ value of 119.0 for 108 dof. 

The possible presence of a fluorescent iron line from neutral material
is suggestive of X--ray irradiation of cold iron close to
the central compact object; however,  the exclusion
from the spectral fitting of the energy range from 15 to 40~keV due to
the nearby contaminating source,
prevents
us to  establish the   presence of another important signature of
reflection, the ``Compton reflection hump".
In fact, the fit with an
exponentially cut-off power-law spectrum reflected from neutral material
({\sc pexrav} model in {\sc xspec}, Magdziarz \& Zdziarski \cite{mz:95})
plus a Gaussian line with energy fixed at 6.4~keV
resulted in an unconstrained reflection scaling factor f$_{\rm refl}$$\approxlt$1.2
(f$_{\rm refl}$=1 for an isotropic source
above an infinite flat disk), a power-law with a photon index of $\sim$1.8
and an iron line with   EW$\approxlt$110~eV.

%-------------------------------------------------------------------------------------------------

\begin{figure}[!ht]
\vskip 0.8truecm
\centerline{\psfig{figure=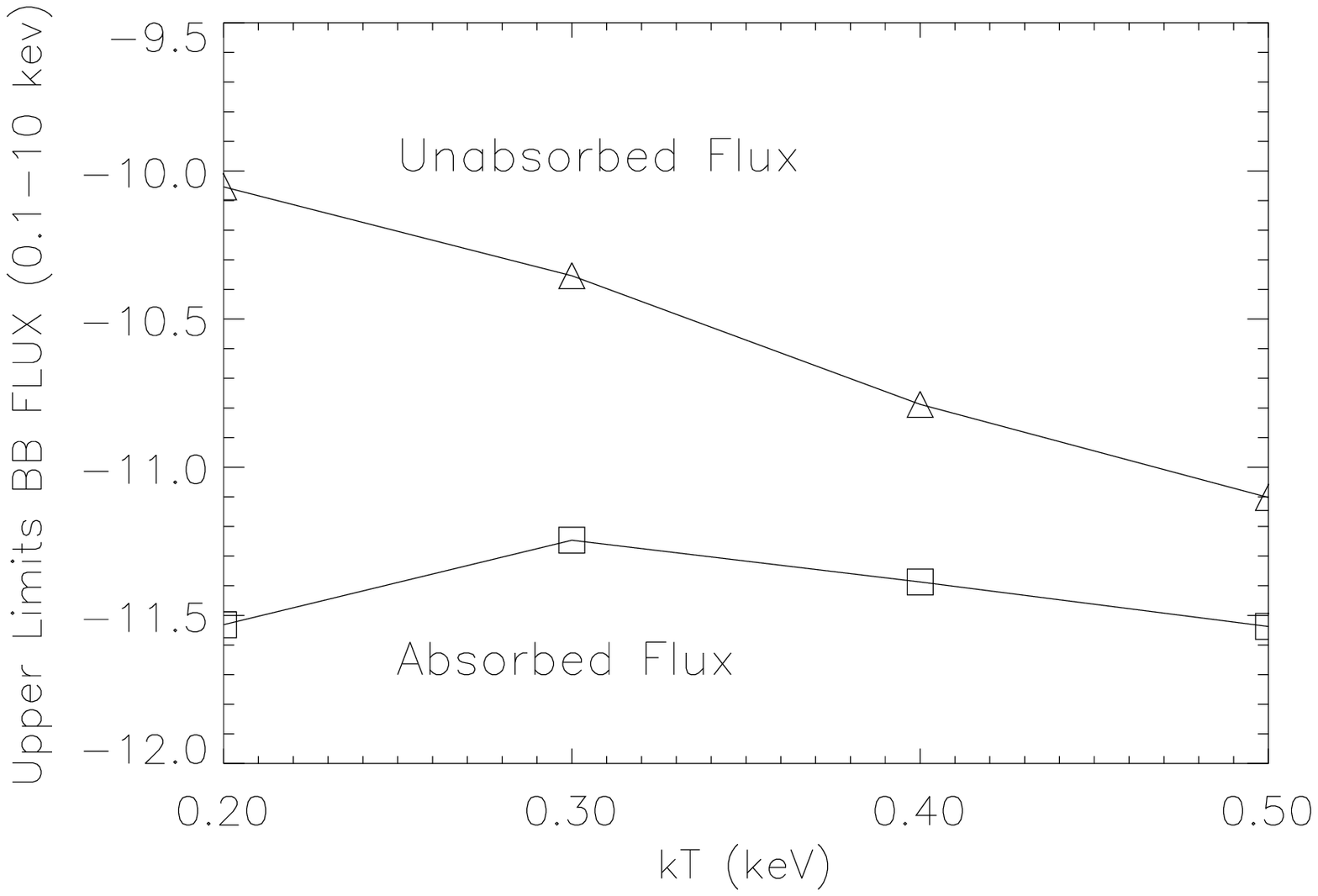,height=80mm,bbllx=70mm,bblly=80mm,bburx=140mm,bbury=250mm}}
\vskip -2.0truecm
\caption{ Upper limits to the presence of a soft component in the \sax\
spectrum of \src. Both the absorbed and unabsorbed   flux
(0.1--10~keV, in logarithmic scale, in units of \ferg) are reported
for a range of assumed blackbody temperatures.
These results have been obtained adopting a broad-band two-component model
composed of a black-body   (with  fixed temperature)   and a power-law
($\Gamma$$\sim$1.66, \nh$\sim$1.85$\times10^{22}$~cm$^{-2}$)
}
\label{fig:bb}
\end{figure}

%-------------------------------------------------------------------------------------------------

We searched  for the possible presence of a soft excess in the \sax\ spectrum,
obtaining a   negative result: adopting a two-component model composed of
a power-law and a blackbody with
temperature in the range 0.2--0.5~keV, we can place the upper limits
shown in Fig.~\ref{fig:bb} to the flux contributed by the blackbody.

\section{Discussion}
\label{sect:discussion}

%-------------------------------------------------------------------------------------------------

\begin{figure*}[!ht]
\vskip 0.0truecm
\centerline{\psfig{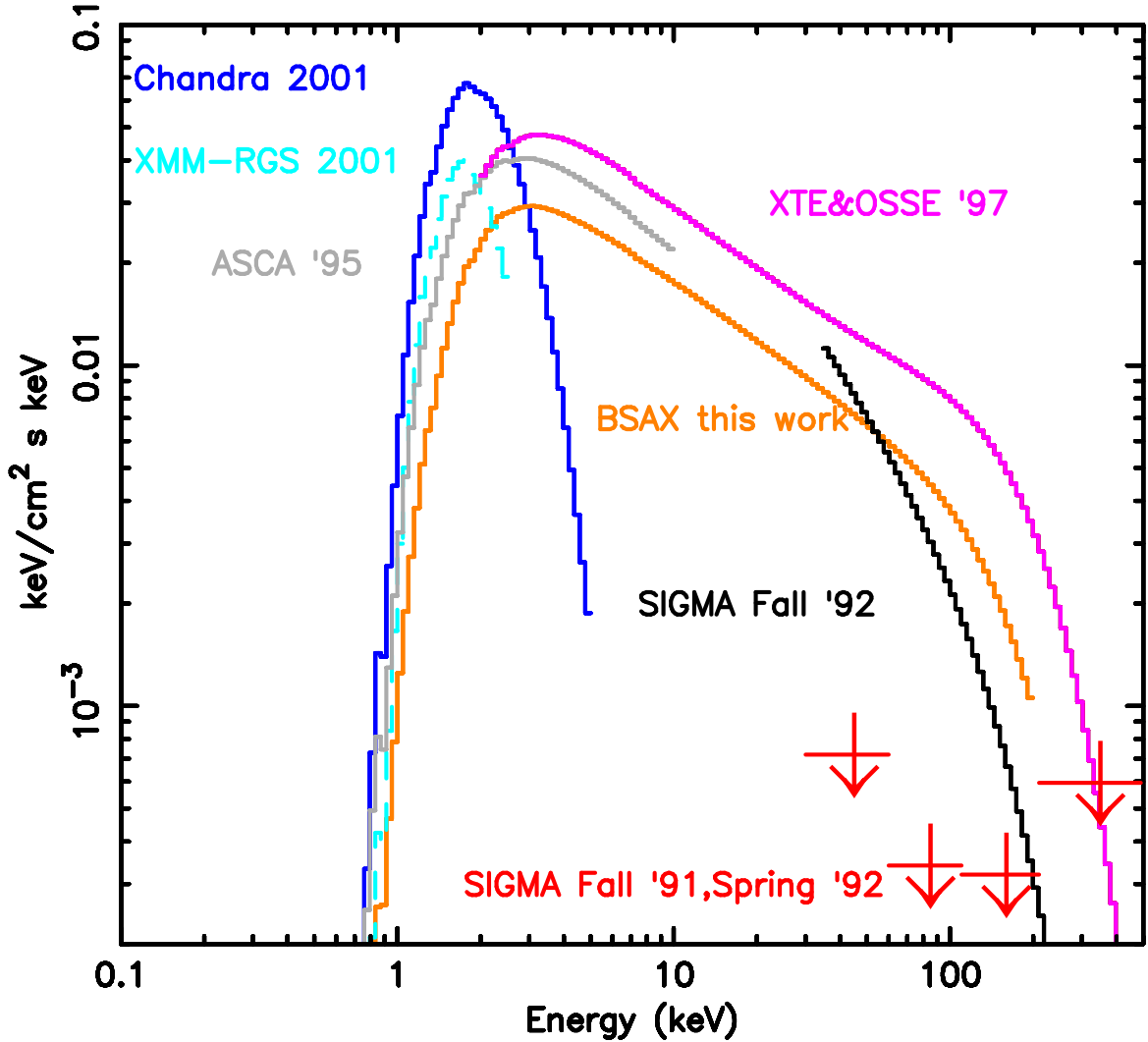}}
\vskip 0.0truecm
\caption{Comparison of representative spectra of
\src. Our \sax\ best-fit is also shown.
The references used to produce this plot have been reported in Table~\ref{tab:ref}
}
\label{fig:cfr}
\end{figure*}

%-------------------------------------------------------------------------------------------------

The  \sax\ observation reported here has allowed us to study the black-hole
candidate \src\ covering simultaneously the broad energy range from 0.1 to 200~keV.
Previously, only Mereghetti et al. (\cite{m:94})
and  Lin et al. (\cite{l:00}) reported broad band spectral studies of \src\ .
Both studies were based on \textit{nearly}  simultaneous observations
with different satellites
and, unfortunately, lacked a complete covering of interesting portions of the spectrum.
Mereghetti et al. (\cite{m:94}) used ROSAT ($<$2~keV) and SIGMA ($>$40~keV) observations,
while Lin  et al. (\cite{l:00}) analyzed multi-wavelength  data
from the radio band to  $\sim$500 keV, but lacked the information for the
soft X--rays below 2 keV, a particularly interesting region in order to establish the
possible presence of a soft spectral component.

%  coming from   the inner regions of the accretion disk.

%--------------------------------------------------------
\begin{table}
\begin{center}
\begin{small}
\caption[]{Summary of the observations used in Fig.~\ref{fig:cfr} }
\begin{tabular}{lll}
\hline
\noalign {\smallskip}
Instrument &   Observation  date  &  Ref. \\
\hline
\noalign {\smallskip}
SIGMA       & Fall 1991-Spring 1992 &  Gilfanov et al. \cite{g:93}  \\
SIGMA       &  Fall 1992            &  Gilfanov et al. \cite{g:93}   \\
ASCA        &  1995 Mar 29          &  Mereghetti et al.  \cite{m:97} \\
XTE \& OSSE &  August  1997         &  Lin et al. \cite{l:00} \\
Chandra     &  2001 Mar 24          &  Heindl \& Smith \cite{hs:01}  \\
XMM-RGS     &  2001 Mar 22          &  Miller et al. \cite{mw:01} \\
\noalign {\smallskip}
\hline
\label{tab:ref}
\end{tabular}
\end{small}
\end{center}
\end{table}
%--------------------------------------------------------------------------

This is a particularly interesting point, in view of previous reports of a
soft spectral component in \src, possibly coming from the inner regions of
the accretion disk. The question is, first, whether such a component is really
present in this source and, second, whether it is associated to a particular
spectral state, as seen in other black hole candidates.

We have collected from the literature all the observations of \src\ and
summarized the main spectral results in Fig. 4.
It appears that most observations found \src\ in a typical  state
characterized by a hard power-law spectrum extending to $\sim$100 keV
followed by a cut-off.
The luminosity in this state is not constant: variability on different timescales
is present (see,e.g., Keck et al. \cite{k:01} for a long term monitoring
with the XTE/ASM and CGRO/BATSE).
In Fig. 4 we have reported two broad band spectra that bracket
the observed range of values: the XTE/OSSE spectrum (Lin et al. \cite{l:00})
and our \sax\ results.

During this spectral state, at least two reliable observations
demonstrate the absence of a soft component:
the one obtained with  ASCA  in March 1995  
(Mereghetti et al. 1997) and  the \sax\ one reported here.
Both missions provided a good coverage of the region around $\sim$1 keV
with imaging instruments that minimized the contamination with the nearby source
GX 5--1.
The possible presence of a soft component suggested by a comparison
of ROSAT PSPC (E$<$2 keV) and SIGMA (E$>$40 keV) data
obtained in 1993 (Mereghetti et al. (1994)) is controversial,
since it was not confirmed by  different analysis of the same data
(Grebenev et al. \cite{g:97},  Keck et al. \cite{k:01}).

Recently (2001, February 27) \src\ entered an ``off'' state,
with a 2--20 keV flux reduced by a factor ten,
smaller time variability and a
spectrum dominated by a thermal
component with  blackbody temperature kT$_{\rm bb}$$\sim$0.4~keV
(Smith et al. \cite{s:01}).
This is  the first clear evidence of a different spectral state,
although it is possible  that \src\ was in a similar  state in 1992--1992 when only an
upper limit could be obtained at energies above 40 keV
(Gilfanov et al. (\cite{g:93}); the upper limits obtained
with SIGMA are indicated in Fig. 4, no  simultaneous observations were available
at lower energies).

An observation during this ``off" state (March 22, 2001)
with the \xmm\ Reflection Grating Spectrometer (RGS)
showed a spectrum dominated by a blackbody component
with kT$_{\rm bb}$$\sim$0.3~keV and with a steep power-law
($\Gamma\sim$2.9)  contributing to
less than 2\% of the total flux in the 0.6--2.3~keV band
(Miller et al. \cite{mw:01}).
Similar results were obtained in an observation performed two days later
with Chandra (Heindl \& Smith\cite{hs:01}).
Preliminary results of an observation performed in September 2000 with
the EPIC instrument on \xmm\ seem to indicate that \src\ was in an ``intermediate''
state, in which both a thermal component and a power-law were required to
fit the 0.2-10 keV spectrum (Goldwurm et al. \cite{g:01}).
In conclusion, it seems that the soft thermal component has been significantly
detected only when \src\ is in a low intensity state.

%%----------- barret box
Keck et al. (\cite{k:01}) have interpreted the long-term behavior of \src\ in term of an
advection-dominated accretion flow model (ADAF),
finding that 1990-1993 nearly simultaneous observations support
the ADAF  predictions.
This interpretation is based on a comparison of the soft (1-20 keV)
and hard (20-200 keV) luminosities (see Barret et al. \cite{bmg:96}).
From our \sax\ observation we can measure for the first time really simultaneously
these quantities,
that, assuming as Keck et al. (\cite{k:01}) a distance of  8.5~kpc,
are $L_{1-20} = 5.4\times10^{36}$~erg~s$^{-1}$ and
 $L_{20-200} =9.7 \times10^{36}$~erg~s$^{-1}$.
 Taking into account the uncertainties of  $\approxlt$10\%,
these luminosity values place \src\
  well inside the so-called
``burster box" in the  $L_{1-20}-L_{20-200}$ plane,
and are not consistent with the ADAF model for a 10.6~$\Msun$ black-hole.
This confirms   that  BHCs in low-state are not spectrally
distinguishable  from neutron stars.

\section{Conclusions}
\label{sect:conclusions}

The \sax\ 0.1--200~keV spectrum of \src\ is adequately
fit with a Comptonized emission model with an electron temperature of
$\sim$30~keV   or,
alternatively, with a power-law with a high energy cut-off at    $\sim$70~keV.
This spectral model  is typical of black-hole candidates in their low/hard state.
The luminosity in the range 0.1--200~keV is 1.4$\times 10^{37}$~erg~s$^{-1}$, and
the measured column density is  almost model-independent in the range
$\sim$$1.8-1.9$$\times10^{22}$~cm$^{-2}$.
 
No soft excess is evident from this observation, and
we can put stringent upper limits
to the flux contributed by a soft component.
A blackbody component at a level similar  to that reported by
Goldwurm et al. (\cite{g:01}),
would have been clearly detected in our \sax\ observation, being
a factor $\sim$7.5 larger than our  upper limits.

We have  marginally   detected (98.3\% confidence level)
a broad line from cold iron (6.4~keV).
The presence of this line  possibly indicates the existence of reflection
from the accretion disk
of hard X--rays coming from the central source.
Another signature of the existence of such a  component
is a  hump in the energy range 20--30 keV (e.g., George \& Fabian \cite{gf:91}).
Unfortunately,
the  X--ray data from  15 to 40 keV  could not be used in  our spectral analysis
due to the contamination from a bright off-axis source.
Thus, the presence of reflection in \src\ deserves further investigations
with imaging instruments covering the hard X--ray region, such as those that
will be soon available with the INTEGRAL satellite.

\end{document}